%% file: main.tex
\documentclass[sigconf,screen,pbalance]{acmart}
\AtBeginDocument{%
  }

\copyrightyear{2026}
\acmYear{2026}
\setcopyright{cc}
\setcctype{by}
\acmConference[FSE Companion '26]{34th ACM Joint European Software Engineering Conference and Symposium on the Foundations of Software Engineering}{July 05--09, 2026}{Montreal, QC, Canada}
\acmBooktitle{34th ACM Joint European Software Engineering Conference and Symposium on the Foundations of Software Engineering (FSE Companion '26), July 05--09, 2026, Montreal, QC, Canada}
\acmDOI{10.1145/3803437.3805571}
\acmISBN{979-8-4007-2636-1/2026/07}

\usepackage{todonotes}
\usepackage{multirow}
\usepackage{enumitem}
\usepackage{caption}
\usepackage[commandnameprefix=always, defaultcolor=red,final]{changes}

\begin{document}
\setlength{\tabcolsep}{3pt}
\title{Projectional Decoding: Towards Semantic-Aware LLM Generation}

\author{Boqi Chen}
\orcid{0000-0002-1451-3603}
\affiliation{%
  \institution{University of Ottawa}
  \city{Ottawa}
  \country{Canada}
}
\email{boqi.chen@uottawa.ca}

\author{Jos\'e Antonio Hern\'andez L\'opez}
\orcid{0000-0003-2439-2136}
\affiliation{%
  \institution{University of Murcia}
  \city{Murcia}
  \country{Spain}}
\email{joseantonio.hernandez6@um.es}

\author{Aren A. Babikian}
\orcid{0000-0002-8108-0043}
\affiliation{%
  \institution{University of Toronto}
  \city{Toronto}
  \country{Canada}
}
\email{babikian@cs.toronto.edu}

\renewcommand{\shortauthors}{Chen et al.}

\newcommand{\aren}[1]{\todo[color=yellow]{#1}}

\begin{abstract}
Large language models (LLMs) are increasingly used to generate software artifacts across many software engineering (SE) tasks, yet ensuring the semantic validity of these artifacts remains a fundamental challenge.
Existing constrained decoding techniques can enforce syntactic correctness and, in some cases, specific semantic rules, but lack a general representation that bridges LLM-generated text with the reasoning required for semantic validation in SE.
In this paper, we propose \emph{projectional decoding}, a novel conceptual framework that integrates domain semantics directly into the generation process by maintaining, alongside text, a partial graph model as the primary artifact representation throughout generation. 
This abstract representation enables incremental semantic validation by explicitly capturing uncertainty and natively supporting error detection, while guiding generation toward semantically valid outputs with provable guarantees. 
We present preliminary results on a program generation task which demonstrate the potential of this approach to improve the semantic validity of LLM-generated artifacts. We also discuss how projectional decoding can enable verifiable automation with LLMs across various SE activities.

\end{abstract}




\maketitle

\input{sections/01.introduction.tex}
\input{sections/02.background.tex}
\input{sections/03.approach.tex}
\input{sections/04.evaluation.tex}
\input{sections/05.conclusion.tex}


\bibliographystyle{ACM-Reference-Format}
\bibliography{biblio.bib}


\end{document}

%% file: sections/01.introduction.tex
\section{Introduction}
\label{sec:introduction}










Large language models (LLMs) are applied across many software engineering (SE) tasks~\cite{zhang2023survey}, including code generation~\cite{huang2023agentcoder,sun2024clover,song2026evaluating}, debugging~\cite{tian2024debugbench}, and software modeling~\cite{lopez2024text2vql,chen2025accurate,di2025use}. Unlike tasks in many other domains, \textit{SE tasks typically require outputs that conform to strict specifications.} Generated artifacts must be both syntactically and semantically valid to satisfy intended requirements. Such validity is essential for ensuring correct behavior~\cite{sun2024clover,ye2025verina}, compliance with regulations~\cite{song2026evaluating,chen2025shieldagent}, and seamless integration with other components~\cite{lyu2025automatic,di2025use}. However, LLMs often struggle to consistently produce syntactically and semantically valid outputs~\cite{willard2023efficient,song2026evaluating,chen2025accurate}. As LLMs are increasingly used in critical SE applications~\cite{gohar2024codefeater}, addressing their semantic validity becomes an immediate task.

Some approaches repair LLM-generated artifacts after complete generation with \textit{iterative post processing}~\cite{he2025llm,chen2025accurate,hong2023metagpt}.
Such post-hoc methods can be inefficient and may not guarantee validity, particularly when the repair process is LLM-based. Invalid outputs may also be non-interpretable, potentially preventing repair techniques from being applied at all. These limitations suggest that semantic validity should also be addressed \textit{during generation}, not only after.

In this paper, we argue that \textit{semantic-aware LLM generation in SE requires shifting from text-focused generation to methods that leverage representations capturing the SE-specific structure of artifacts}.
While LLMs generate text, SE artifacts are inherently structural and often represented using graph-based models~\cite{famelis2012partial,marussy2020specification}. Moreover, many semantic properties in SE are specified as constraints over such models~\cite{richters2002ocl,bergmann2010incremental,pnueli1977temporal}. Treating semantic properties at the text level therefore introduces \textit{a fundamental mismatch between how artifacts are generated and how their correctness is defined}.

Recent work on constrained decoding partially addresses this gap by guiding LLM generation using syntactic constraints~\cite{beurer2023prompting,llguidance2024,willard2023efficient,ugare2024syncode,dong2025xgrammar}, and in some cases lightweight semantic rules~\cite{mundler2025type,ugare2025itergen,ma2025logically,nagy2026chopchop}.
However, most existing methods still treat LLM outputs primarily as text,
resulting in insufficient structural expressiveness and limited generalization to a broader range of SE scenarios.

We argue that \textit{bridging this representation gap requires maintaining an explicit abstract representation during LLM generation}. Thus, we propose \emph{projectional decoding}, a model-based constrained decoding framework inspired by the concept of projectional editing~\cite{voelter2016efficient,berger2016efficiency,lafontant2020gentleman}, where a primary model representation is maintained during editing to enable continuous validation and early error detection. Similarly, projectional decoding maintains a \emph{partial model}~\cite{famelis2012partial} of the artifact during generation that captures both already generated structures and the uncertainty of yet-to-be-generated elements. 

During generation, each new token refines the partial model.
This refinement enables \textit{incremental semantic validation directly at the model level} and allows the integration of established model-based reasoning techniques~\cite{semerath2017graph,cousot1977abstract}, supporting generation strategies that better capture SE-specific nuances than text-first approaches.


We introduce projectional decoding as \textit{guided generation approach}, using it to derive semantic guardrails for the LLM token sampling process.
We present a preliminary feasibility evaluation on a CLEVR program generation task~\cite{johnson2017clevr}.
The results show that even a simple token-masking strategy informed by partial models can substantially improve semantic validity and task accuracy compared to standard decoding or syntax-only constrained decoding.

%% file: sections/02.background.tex
\section{Background and Challenges}
\label{sec:background}

\textbf{Semantics in SE artifacts.}
Semantics captures the meaning and intended properties of software artifacts beyond their syntactic structure.
In SE applications, model-based methods are commonly used to define graph-based semantics that capture the complex interconnections within a domain.
Engineers formalize semantics by defining \textit{a domain metamodel together with well-formedness constraints}. Such metamodel can be linked explicitly with the grammar rules with existing tools~\cite{jetbrains_mps,bettini2016implementing}.
Constraints specify properties that metamodel instances (representing software artifacts) must satisfy and are often expressed using graph patterns~\cite{richters2002ocl,bergmann2010incremental} or temporal logic~\cite{pnueli1977temporal}.
Prior work has considered various semantic constraints for LLM-generated artifacts, including invariants~\cite{sun2024clover}, behavioral specifications~\cite{song2026evaluating,chen2025shieldagent}, and structural consistency~\cite{lopez2024text2vql,chen2025accurate}. 
Existing studies consistently show that LLMs are prone to violating such constraints, 
even under explicit prompting~\cite{chen2025accurate,chen2025shieldagent,lopez2024text2vql}.


\noindent
\textbf{Constrained decoding} refers to a class of approaches for producing structured outputs by modifying the LLM generation strategy. Existing methods primarily focus on \textit{enforcing syntactic correctness according to predefined grammar rules}, typically context-free grammars~\cite{beurer2023prompting,llguidance2024,willard2023efficient,ugare2024syncode,dong2025xgrammar}. In general, these approaches guide generation by masking invalid tokens during sampling from the output distribution~\cite{beurer2023prompting,willard2023efficient}.
With recent advances in optimization techniques, the computational overhead of constrained decoding has been substantially reduced, even negligible in some cases~\cite{llguidance2024}.

More recently, some approaches extend constrained decoding beyond syntax to address semantic validity.
However, these methods typically target particular classes of semantic constraints~\cite{mundler2025type,ma2025logically} and rely on checking generated text directly rather than reasoning over the underlying structure~\cite{ugare2025itergen}.
ChopChop~\cite{nagy2026chopchop} \chadded{reasons over AST spaces to enable semantic pruning at the structural level. Projectional decoding targets more general SE artifacts whose semantics are defined over typed graphs rather than trees, explicitly representing uncertainty through partial models to support incremental validation on general constraints using graph patterns.}

\noindent
\textbf{Open challenges.}
We identify three key challenges in closing the representation gap for semantic-aware LLM generation:

\noindent
\textit{(\textbf{C1}) Abstract representation for semantic evaluation.} Ensuring semantic validity requires an expressive representation to enable automated error detection \textit{during} the generation of SE artifacts.


\noindent
\textit{(\textbf{C2}) Capturing uncertainty.} During generation, an LLM produces only an incomplete artifact, while the remaining is yet to be generated. Evaluating semantic constraints over incomplete artifacts requires explicit consideration of \textit{uncertainty}.


\noindent
\textit{(\textbf{C3}) Effective semantic guidance.} LLM generation is a search process over tokens. Efficiently guiding generation to a semantically valid solution with guarantees requires \textit{reasoning} under uncertainty to enable effective exploration of the search space.

%% file: sections/03.approach.tex
\section{Towards Semantic-Aware LLM Generation}
\label{sec:approach}

\begin{figure}
    \centering
    \includegraphics[width=0.9\linewidth]{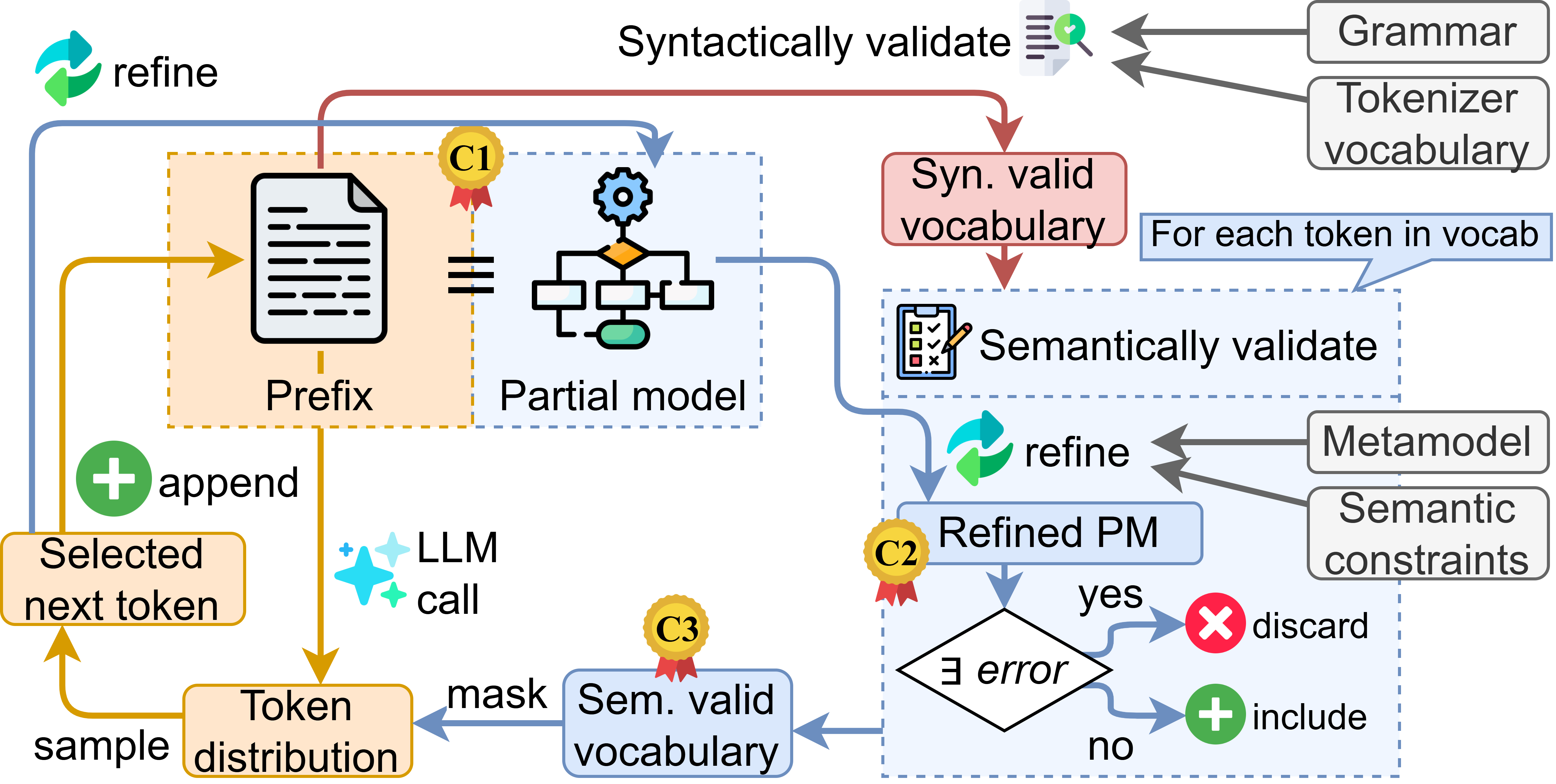}
    \caption{Architecture of projectional decoding}
    \label{fig:architecture}
\end{figure}

To address the challenges and enable semantic-aware LLM generation, we propose \textit{projectional decoding}, a framework for constrained decoding that explicitly integrates structured representations into the LLM generation process.
\autoref{fig:architecture} illustrates the architecture of our proposed projectional decoding framework when applied to guided LLM generation. In the figure, gray nodes denote user-provided inputs, orange represents the standard LLM generation pipeline, red components correspond to syntax-guided generation used in existing approaches, and blue denotes the proposed semantic-guided generation enabled by projectional decoding.

Inspired by \textit{projectional editing} techniques, projectional decoding maintains a partial model representation~\cite{famelis2012partial,marussy2020specification} equivalent to the text artifact throughout generation. This partial model serves as the primary representation for continuously validating semantic constraints as decoding progresses, thereby addressing \textbf{C1}.


At each decoding step, the LLM produces a next token probability distribution over vocabulary based on the current output prefix. In parallel, the vocabulary is filtered by two validation modules. First, \textit{syntactic validation} removes tokens that violate the grammar. Next, \textit{semantic validation} evaluates each remaining token by refining a partial model and checking for any constraint violations. This refinement captures uncertainty in the generation by introducing or updating elements in the partial model according to the metamodel and constraints, addressing \textbf{C2}. Tokens whose refinements lead to semantic violations are discarded,
while the remaining tokens are used to mask the LLM output distribution for effective guidance, addressing \textbf{C3}. \chadded{In practice, most tokens are discarded by the syntactic validation, 
making semantic validaiton computationally tractable even without additional caching or pruning}. The selected token is then appended to the output prefix and is used to refine the partial model. The process repeats until the generation is terminated.

\noindent
\textbf{Abstract representation.}
\textit{The construction and maintenance of a partial model} enables automated violation detection through explicit support for \textit{error states} during generation.
This partial model serves as the primary enabler of semantic validation through its direct connection to the metamodel and constraints, which is not possible when text is used as the primary artifact.


Partial models were originally introduced to capture uncertainty during modeling processes~\cite{famelis2012partial} and can be represented with a specification language~\cite{marussy2020specification}. In projectional decoding, the partial model represents an incomplete instance of a metamodel using a \textit{graph structure}. Each element (node and edges) corresponds to a metamodel element and is associated with a \textit{four-valued logic}~\cite{marussy2020specification}: \textit{certain} elements correspond to parts of the output that have already been completed, \textit{possible} elements are still to be generated in subsequent steps, \textit{absent} elements are ruled out during decoding according to the constraints, therefore excluded from the partial model, and \textit{error} elements indicating violations of semantic constraints.

While \textit{certain} elements can be derived directly from the already generated output, \textit{absent} and candidate \textit{possible} elements are inferred from the syntactic and semantic constraints based on the incomplete portions of the output. This inference process is analogous to automatic code completion in projectional editors~\cite{voelter2016efficient}. \textit{Error} elements may be introduced as a result of validation when no refinement of the partial model satisfies the input constraints.



\noindent\textbf{Capturing uncertainty.}
A key feature of partial models is their ability to represent uncertainty. This features allows syntactic and semantic constraints to be evaluated directly at each generation step, before and after refinement, which enables reasoning about potential new elements or any resulting violations.


\textit{Syntactic validation.}
Based on the grammar and output prefix, various existing constrained decoding techniques~\cite{llguidance2024,willard2023efficient} can be used to determine the set of syntactically valid tokens, which ensure the syntactic correctness of the underlying partial model.


\textit{Refinement and semantic validation.}
For each syntactically valid token, the partial model is updated to reflect the revised state of the generated artifact. This update consists of two main steps: (1) refining existing partial elements based on the newly generated token, analogous to updating auto-completion suggestions, and (2) introducing new (partial) elements as required by the metamodel and semantic constraints. This update process can be realized using existing partial model refinement techniques~\cite{varro2018towards}. 

During refinement, the status of existing elements may change from \textit{possible} to either \textit{certain} or \textit{absent}, depending on whether they are consistent with the provided token. When the refinement of an element violates any constraint, the element will be labeled as \textit{error}. This consistency check can be implemented using existing graph pattern matching techniques~\cite{semerath2017graph,bergmann2010incremental} for structural constraints, or abstract interpretation techniques~\cite{cousot1977abstract} for behavior constraints.

Any model containing \textit{error} elements is considered invalid and discarded.
Tokens that do not introduce violations preserve local validity and ensure that a valid output may potentially be generated.

\noindent\textbf{Effective guided generation.}
The partial model not only captures the uncertainty introduced during generation but also constrains the search space by \textit{reasoning about possible future valid elements}.
This enables the integration of various guided generation strategies. A naïve approach is to sample the next token from the LLM's distribution restricted to tokens that satisfy both syntactic and semantic constraints. The sampled token is then appended to the prefix and used to refine the partial model for the next generation step. Generation proceeds until termination when either a termination token is produced or no semantically valid refinements remain. We discuss the potential of more advanced strategies in \autoref{sec:future_plan}.

\begin{figure}[tb]
    \centering
    \includegraphics[width=0.9\linewidth]{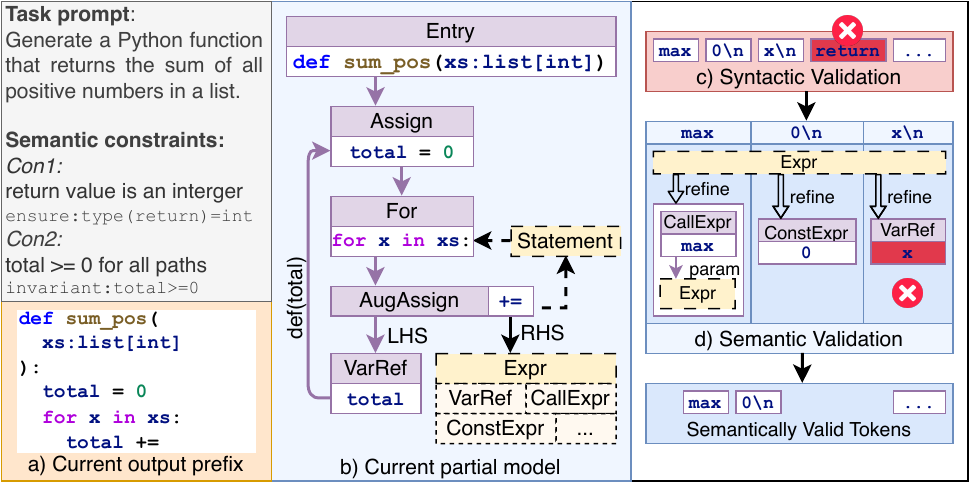}
    \caption{An example of projectional decoding for code generation; some partial model elements are excluded for clarity}
    \label{fig:example}
\end{figure}

\noindent\textbf{Example.}
\autoref{fig:example} shows an example of projectional decoding applied to code generation. The task prompt specifies the intended behavior of a Python function 
and includes two semantic constraints (defined in a dedicated constraint language) specifying a postcondition (\textit{Con1}) and invariant (\textit{Con2}). Such constraints commonly arise in SE settings such as contract-based code generation~\cite{sun2024clover}.

The grammar rules define the Python syntax for parsing.
In our approach, the LLM incrementally generates a Python program.
An example of a partially syntactically valid Python program is shown in \autoref{fig:example}.a.
In parallel, our method extracts structural information (e.g., abstract syntax, control flow, and data flow) and incrementally constructs a corresponding partial model, shown in \autoref{fig:example}.b.
In this partial model, \textit{certain} elements (in purple) correspond to already-generated program components, while \textit{possible} elements (in yellow) represent incomplete parts inferred from the metamodel and constraints. For example, the \texttt{Expr} element represents the incomplete right hand side of the \texttt{AugAssign} statement.

Given the output prefix, syntactic validation derives the set of syntactically valid tokens shown in \autoref{fig:example}.c. For example, \texttt{return} is invalid on the right hand side of the \texttt{AugAssign} and is excluded. Each remaining token is then evaluated by semantic validation (\autoref{fig:example}.d) through partial model refinement. Some tokens, such as \texttt{max}, introduce new \textit{possible} elements,
while others, such as \texttt{x\textbackslash n} and \texttt{0\textbackslash n}, only refine existing ones.
Abstract interpretation~\cite{cousot1977abstract} on the model can identify that \texttt{x\textbackslash n} may cause the value of \texttt{total} to be negative and therefore violates constraint \textit{Con2}, leaving only \texttt{max} and \texttt{0\textbackslash n} as candidates for the final sampling step. These valid tokens will then be used to mask the LLM-generated token distribution.




%% file: sections/04.evaluation.tex
\section{Preliminary Evaluation}
\label{sec:evaluation}

We conduct a preliminary evaluation on the CLEVR program generation task~\cite{johnson2017clevr} to show the feasibility of projectional decoding, involving generating programs based on a question description. It is originally designed for visual question answering but has also been used as a challenging benchmark for LLM-based generation~\cite{chen2025accurate}.

\begin{figure}[tb]
  \centering
  \includegraphics[width=0.9\linewidth]{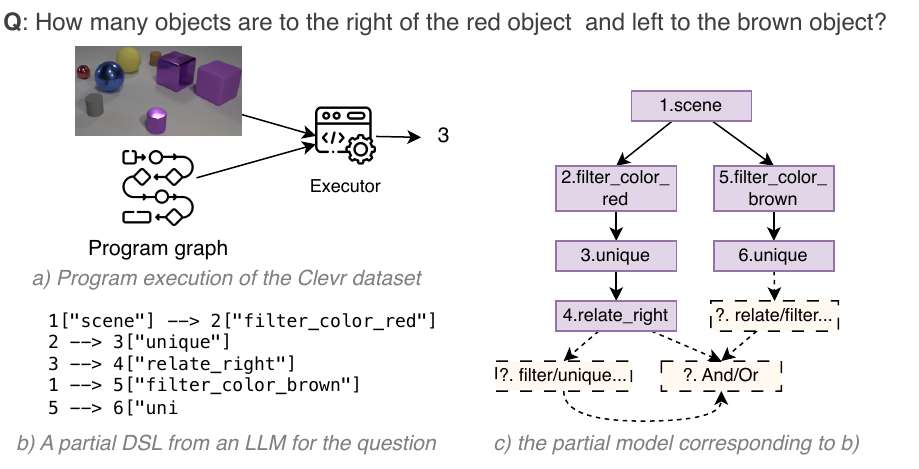}
  \caption{An example of the CLEVR program}
  \label{fig:clevr_example}
\end{figure}

\autoref{fig:clevr_example} shows an example of CLEVR program generation. The goal is to generate a program in a domain-specific language (DSL) that computes the answer to a question with a scene graph. The generated program can be evaluated by execution to obtain an answer (\autoref{fig:clevr_example}.a) and compare it with the ground truth answer. We reuse existing semantic constrains~\cite{chen2025accurate} to ensure the generated program is executable, including (Con1) type consistency (e.g., the argument of \texttt{filter} must be of type \texttt{set}), (Con2) number of inputs (e.g., \texttt{And} must have exactly two inputs), and (Con3) acyclic structure, etc.

\noindent\textbf{LLMs.}
We evaluate the approaches with three Qwen3~\cite{yang2025qwen3} variants, including 4B, 8B, and 14B parameters. Qwen3 is an open-source LLM that has showed strong performance on various generation tasks. While our approach is agnostic to the LLM, closed-source LLMs are not evaluated due to the lack of access to their internal token distribution required for constrained decoding. To ensure a consistent evaluation, we use greedy sampling (temperature=0) for all approaches with a few-shot prompt containing four examples.

\noindent\textbf{Compared approaches.}
We compare three approaches: (1) \textit{No Guidance}, where the LLM generates the program without any guidance; (2) \textit{Syntactic}, where syntactic constrained decoding is applied, we use LLGuidance~\cite{llguidance2024}; and (3) \textit{Semantic}, where projectional decoding is used.  \autoref{fig:clevr_example}.b and c show a partial CLEVR DSL and the maintained partial model. In this case, several partial elements (e.g., \texttt{filter}, \texttt{And}, and \texttt{relate}) are derived by reasoning using the semantic constraints. For example, the \texttt{unique} operator must be followed by a \texttt{filter} operator according to the type constraint and the two branches must be eventually joined by an \texttt{And} or \texttt{Or} operator since there can only be a single output node.

\noindent\textbf{Metrics.}
We evaluate the approaches based on several metrics: (1) \textit{Syntactic validity (Syn.)}, the percentage of syntactically correct programs; (2) \textit{Semantic validity (Sem.)}, the percentage of programs satisfying all semantic constraints; and (3) \textit{Accuracy (Acc.)}, the percentage of programs producing the correct answer when executed on a scene graph. All metrics are computed over all generated results. For semantic guidance, we also measure the average generation time per token (\textit{T}) relative to the no-guidance approach.

\begin{table}[tb]
    \centering
    \caption{Preliminary evaluation results on CLEVR program generation with semantic constraints for Qwen3 (in \%).}
    \label{tab:preliminary_results}
  \scalebox{0.85}{
    \begin{tabular}{l|ccc|ccc|cccc}
        \toprule
        \multirow{2}{*}{\textbf{Size}} & \multicolumn{3}{c|}{\textbf{No Guidance}} & \multicolumn{3}{c|}{\textbf{Syntactic}} & \multicolumn{4}{c}{\textbf{Semantic (Ours)}} \\
        & \textbf{Syn.} & \textbf{Sem.} & \textbf{Acc.} & \textbf{Syn.} & \textbf{Sem.} & \textbf{Acc.} & \textbf{Syn.} & \textbf{Sem.} & \textbf{Acc.} & \textbf{T} \\
        \midrule
        4b & 9.00 & 4.33 & 2.33 & 100 & 48.67 & 28.00  & 99.33 & \textbf{73.67} & \textbf{36.00} & 1.1x \\
        \midrule
        8b & 87.67 & 60.33 & 38.67 & 100 & 61.00 & 37.33 & 99.00 & \textbf{79.67} & \textbf{40.00} & 1.5x \\
        \midrule
        14b & 83.33 & 55.44 & \textbf{37.67} & 100 & 58.33 & 36.67 & 99.67 & \textbf{73.33} & 37.33 & 1.1x \\
        \bottomrule
    \end{tabular}
  }
\end{table}

\noindent\textbf{Results.}
As shown in \autoref{tab:preliminary_results}, projectional decoding significantly improves semantic validity across all LLMs compared to both baselines. Notably, even the worst semantic validity achieved by projectional decoding (73.33\% with 14B LLM) substantially outperforms the best semantic validity of other approaches (61.00\% with syntactic decoding on 8B LLM). However, none of the LLMs reach a perfect semantic validity even with semantic guidance. This result may be caused by the limitation that LLM generation explored a decoding path where no semantically valid solutions are possible, \chadded{in which case generation terminates early with a semantically invalid output}. Future work can explore advanced search strategies such as backtracking to further improve the semantic validity.

The improvement for semantic validity also leads to better task accuracy in 2 out of 3 cases, achieving the best overall accuracy on the 8B LLM (40\%), demonstrating the practical benefits of enforcing semantic constraints during generation. The syntactic guidance only provides improvements on accuracy for the smallest 4B LLM, showing that syntactic correctness alone is insufficient to ensure high-quality generation. Moreover, the time overhead introduced by semantic guidance is moderate, ranging from 1.11x to 1.50x across different LLMs compared to no guidance.



%% file: sections/05.conclusion.tex
\section{Applications and Research Directions}
\label{sec:future_plan}
\noindent
\textbf{Application.}
Projectional decoding can be used to guarantee semantic validity in many SE applications, particularly in design-by-contract scenarios. As shown in \autoref{sec:approach} and \autoref{sec:evaluation}, it can be applied to contract-based and DSL-based program generation with invariants and structural constraints. This form of constraint-guided generation naturally extends to other settings, including compliance with safety regulations in API calling~\cite{song2026evaluating,chen2025shieldagent}, enforcing well-formedness constraints in software modeling~\cite{chen2025accurate}, and maintaining cross-artifact consistency in larger software projects~\cite{sultan2024ai}.

More broadly, by integrating a abstract semantic representation, projectional decoding supports \textit{decoding-time validation of several interdependent artifact generation} at the same time, where generation of one artifact (e.g., tests, design, or safety cases) immediately constrains and informs the generation of others (e.g., code). This co-generation capability shows its potential as a unifying framework for semantic-aware automation with LLMs across the SE lifecycle.

\noindent
\textbf{Research directions.}
Realizing the broader vision with projectional decoding raises several research directions for investigation.

\textit{Effectiveness.}
By maintaining a partial model, projectional decoding enables more advanced search strategies than text-only decoding and supports both token-level~\cite{ugare2025itergen} and model-level backtracking during the generation process. Uncertainty-aware heuristics, such as prioritizing models with fewer \textit{possible} elements, could further improve the effectiveness of semantic guidance~\cite{kapoor2025constrained}.

\textit{Efficiency.}
Although the overhead observed in the preliminary evaluation is modest, further optimizations are possible. For example, the validation could be parallelized with LLM inference, and incremental constraint evaluation techniques~\cite{bergmann2010incremental} could be explored to reduce redundant computations in decoding steps.

\textit{Generalization.}
Generalizing projectional decoding beyond specific use cases requires systematic support for partial modeling using a generalized framework. Existing work on partial model specification~\cite{marussy2020specification} and reasoning~\cite{marussy2024refinery} provides a promising foundation for defining reusable abstractions and constraints that could enable projectional decoding across various SE domains.



\section{Conclusion}
\label{sec:conclusion}

This paper proposes \emph{projectional decoding}, a novel conceptual framework for constrained decoding addressing the representation gap for semantic evaluation by maintaining a partial model during generation. Projectional decoding tackles key challenges by providing an abstract representation for semantic evaluation, capturing uncertainty during generation, and enabling effective semantic guidance. To realize the vision of semantic-aware LLM generation, future work will explore advanced search strategies, optimizations, and broader applications in SE tasks requiring semantic validity in comparison with existing structured decoding approaches.

